\documentclass[fancybox,11pt,letterpaper]{article}
\usepackage{epsfig,graphicx,color}
\usepackage[cp850]{inputenc}
\usepackage[T1]{fontenc}
\usepackage[english]{babel}
\usepackage{cite}
\usepackage{amssymb}
\usepackage{rotating}
\usepackage{amsmath}
\usepackage[f]{esvect}

\def\tablenotes{\bgroup\parfillskip=0pt plus 1fil
\leftskip=0pt\relax \rightskip=0pt
\vskip2pt\footnotesize}
\def\endtablenotes{\vskip1pt\egroup}

\def\sphline{\noalign{\vskip3pt}\hline\noalign{\vskip3pt}}

\allowdisplaybreaks

\begin{document}

\title{One-Center Nonrelativistic Integrals of Second Order for the NMR Shielding Tensor}

\author{B. Abouzaid$^{\dag}$, M. Essaouini$^{\ddag}$ and H. Safouhi$^{\S}$\footnote{The corresponding author (HS) acknowledges the financial support from the Natural Sciences and Engineering Research Council of Canada~(NSERC) - Grant RGPIN-2016-04317}\\
\\
$^{\dag}${\it \'Ecole Nationale des Sciences Appliqu\'ees}\\
{\it Laboratoire des sciences de l'ing\'enieur pour l'\'energie}\\
{\it El Jadida, MA 24002}\\
{\it bhabouzaid@yahoo.fr}\\
\\
$^{\ddag}${\it Department of Mathematics, Faculty of Sciences}\\
{\it Chouaib Doukkali University, El Jadida, MA}\\
{\it essaouini.m@ucd.ac.ma}\\
\\
$^{\S}${\it Mathematical Division}\\
{\it Campus Saint-Jean, University of Alberta}\\
{\it 8406  91 Street, Edmonton (AB) T6C 4G9, Canada}\\
{\it Edmonton (AB), Canada}\\
{\it hsafouhi@ualberta.ca}}

\date{}

\maketitle

{\bf \large Abstract}. \hskip 0.15cm

This work presents an analytical development for one-center nonrelativistic integrals of second order for the nuclear magnetic resonance (NMR) shielding tensor. The main difficulty in the treatment of these integrals arises from the presence of $r^{-3}$ in the operator. Compact analytical formulae are obtained using $B$ functions as the basis set of atomic orbitals, the Fourier transform formalism and Cauchy's residue theorem. The obtained formulae are computationally convenient and can be computed to machine accuracy.

\vspace*{0.5cm}

Nuclear magnetic resonance; Shielding tensor; Hamiltonian terms of second orders; $B$ functions; Fourier transform.

\clearpage
\section{Introduction}

Experimental methods based on magnetic resonance are among the most used techniques for investigating molecular and electronic structure. Nuclear magnetic resonance (NMR) parameters are of great interest in chemistry, biology and solid-state physics and their computation for any of the standard models of quantum chemistry constitute a significant challenge~\cite{Vaara-5399-07, Helgaker-Jaszunski-Ruud-99-293-99, Fukui-Baba-Shiraishi-Imanishi-Kudo-Mori-Shimoji-102-641-04, London-8-397-37, Ditchfield-27-789-74, Pyykko-88-563-88, Pople-McIver-Ostlund-49-2960-68, Ishida-24-1874-03, Dickson-Ziegler-100-5286-96, Schreckenbach-Ziegler-99-606-95, Watson-Handy-Cohen-Helgaker-120-7252-04, Autschbach-Ziegler-9-306-02, Kaupp-Buhl-Malkin-04}. Calculations involving a magnetic field should preserve gauge invariance. This is conveniently accomplished by using gauge including atomic orbitals (GIAOs)~\cite{London-8-397-37, Pyykko-88-563-88, Wolinski-Hinton-Pulay-112-8251-90}, constructed using atom-centered basis functions with explicit field dependence.

In ab initio calculations, each molecular orbital (MO) is built from a linear combination of atomic orbitals (LCAO). Thus, the choice of reliable basis functions is of primary importance. Magnetic properties are sensitive to the quality of the basis set due to many contributing physical phenomena arising from both the vicinity of the nucleus and from the valence region. For this reason, it is highly desirable to use exponential type functions (ETFs) which are better suited than Gaussian type functions (GTFs)~\cite{Boys-200-542-50, Boys-201-125-50} to represent electron wave functions near the nucleus and at long range. Among ETFs, Slater type functions (STFs)~\cite{Slater-42-33-32} and $B$ functions~\cite{Filter-Steinborn-18-1-78} are undoubtedly the most popular. ETFs decay exponentially for large distances~\cite{Agmon-85} and satisfy Kato's conditions for exact solutions of the appropriate Schr\"{o}dinger equation~\cite{Kato-10-151-57}.  Although interest in using ETFs in the computation of NMR parameters is increasing~\cite{Schreckenbach-Ziegler-99-606-95, Dickson-Ziegler-100-5286-96, Watson-Handy-Cohen-Helgaker-120-7252-04}, no effort has yet been dedicated to their analytical treatment over ETFs. Straightforward numerical integration was used for the computation of integrals associated with these parameters.

Of the NMR parameters, the nuclear shielding tensor is of utmost importance. The main difficulty in the calculation of the shielding tensor arises from the operators associated with these parameters which lead to extremely complicated integrals that are not present in the usual ab initio Hartree-Fock calculations. An example of such operators is $\frac{(\vec{u}_\alpha \times \vec{r}_{j\nu} )\cdot(\vec{u}_\beta \times \vec{r}_{jN})}{r_{jN}^3}$, where $\vec{u}_\alpha$ and $\vec{u}_\beta$  are unitary vectors of the cartesian referential, and $\vec{r}_{jN}$ and $\vec{r}_{j\nu}$ are the vectors separating the $j^{\textrm{th}}$ electron from the $N^{\textrm{th}}$ and the $\nu^{\textrm{th}}$  nuclei respectively. The finite-perturbation method~\cite{Pople-McIver-Ostlund-49-2960-68} can be used to compute the NMR parameters~\cite{Malkina-Salahub-Malkin-105-8793-96}, but the numerical differentiation can be very unstable and this is why analytical development has to be used in such calculations~\cite{Safouhi33, Safouhi39, Safouhi41}.

In~\cite{Safouhi33}, we used properties of unormalized STFs and $B$ functions, along with the Fourier transform method~\cite{Trivedi-Steinborn-27-670-83, Grotendorst-Steinborn-38-3875-88} to derive analytical formulae for three-center nuclear shielding tensor integrals. These analytical formulae involve semi-infinite spherical Bessel integrals have proven to be a computational challenge.

The present contribution pertains to the analytical development of the one-center case of the shielding tensor integrals. The proposed approach uses $B$ functions as a basis of atomic orbitals, which are better suited for the Fourier transform formalism than STFs. This leads to considerable simplifications in the calculation.  The obtained formulae involve semi-infinite integrals which we were able to solve analytically using Cauchy's residue theorem.  The analytical formulae derived for the one-center shielding tensor integrals does not require any numerical integration and can be computed to machine accuracy.

\section{General definitions and properties}
The functions  $B_{n,l}^m(\zeta,\vec{r})$ are defined by~\cite{Filter-Steinborn-18-1-78}:
\begin{equation}
B_{n,l}^m(\zeta,\vec{r}) = \frac{(\zeta r)^l}{2^{n+l}(n+l)!} \, \hat{k}_{n-\frac{1}{2}}(\zeta r) \, Y_l^m(\theta_{\vec{r}},\varphi_{\vec{r}}),
\end{equation}
where $n$, $l$, and $m$ are the quantum numbers and $\hat{k}_{n-\frac{1}{2}}(z)$ stands for the reduced spherical Bessel function of the second kind and is given by~\cite{Shavitt-63, Steinborn-Filter-38-273-75}:
\begin{eqnarray}
\hat{k}_{n+\frac{1}{2}}(z) & = & z^n \, e^{-z} \, \sum_{j=0}^{n} \frac{(n+j)!}{j! \,(n-j)!} \frac{1}{(2\,z)^{j}}
\label{EQREBE} \\ & = & \sqrt{\frac{2}{\pi}} z^{n+\frac{1}{2}} K_{n+\frac{1}{2}}(z),
\end{eqnarray}
where $K_{n+\frac{1}{2}}(z)$ is the modified Bessel function of the second kind of order $n+\frac{1}{2}$~\cite{Safouhi37}.

The surface spherical harmonic $Y_l^m(\theta_{\vec{r}},\varphi_{\vec{r}})$ is defined explicitly using the Condon-Shortley phase convention for non-negative values of $m$ as follows~\cite{Bransden-Joachain-00, Condon-Shortley-51}:
\begin{equation}
Y_{l}^{m}(\theta_{\vec{r}},\varphi_{\vec{r}}) \,=\, (-1)^m
\left[\frac{(2l+1)(l-m)!}{4 \pi (l+m)!}\right]^{\frac{1}{2}}
P_l^m\left(\cos(\theta_{\vec{r}})\right) \, \mathrm{e}^{i \, m \, \varphi_{\vec{r}}},
\label{EQSPHA}
\end{equation}
where $P_{l}^{m}(x)$ is the associated Legendre polynomial of $l^{\textrm{th}}$ degree and $m^{\textrm{th}}$ order.

Unormalized Slater type functions (STFs) are defined by~\cite{Slater-42-33-32}:
\begin{equation}
\chi_{n,l}^{m}(\zeta,\vec{r}) \;=\;  r^{n-1} \, e^{-\zeta r} \, Y_{l}^{m}(\theta_{\vec{r}},\varphi_{\vec{r}}).
\label{eq:STOu}
\end{equation}

Unormalized STFs can be expressed as finite linear combinations of $B$ functions~\cite{Filter-Steinborn-18-1-78}:
\begin{equation}
\chi_{n,l}^{m}(\zeta,\vec{r})\;=\; \frac{1}{\zeta^{n-1}} \; \sum_{p=\tilde{p}}^{n-l} \,
\frac{(-1)^{n-l-p} \,\, 2^{2 p + 2 l - n} \,\, (l+p)!}{(2p-n+l)! \,\, (n-l-p)!}\, B_{p,l}^{m}(\zeta,\vec{r}),
\label{eq:STOonB}
\end{equation}
where:
\begin{equation}
  \tilde{p} =
  \left\lbrace
    \begin{array}{llll}
      \displaystyle \frac{n-l}{2}   \quad & \textrm{if} & \quad n-l \quad  & \textrm{is even}\\[0.25cm]
      \displaystyle \frac{n-l+1}{2} \quad & \textrm{if} & \quad n-l \quad  & \textrm{is odd}.
    \end{array}
  \right.
  \label{eq:ptilde}
\end{equation}

Gaunt coefficients are defined by~\cite{Gaunt-228-151-29, Weniger-Steinborn-25-149-82}:
\begin{equation}
\left< l_1 m_1|l_2 m_2|l_3 m_3 \right> = \int_{0}^{2\pi}\int_{0}^{\pi} [Y_{l_1}^{m_1}(\theta,\varphi)]^{*} \, Y_{l_2}^{m_2}(\theta,\varphi) \, Y_{l_3}^{m_3}(\theta,\varphi) \, \sin(\theta){\rm\,d}\theta{\rm\,d}\varphi.
\label{Gaunt}
\end{equation}

The Gaunt coefficients linearize the product of two spherical harmonics:
\begin{equation}
\left[Y_{l_1}^{m_1}(\theta,\varphi)\right]^* Y_{l_2}^{m_2}(\theta,\varphi) \,=\,  \sum_{l=l_{\min},2}^{l_{1}+l_{2}}  \left< l_2\, m_2 | l_1\, m_1 | l\, m_2-m_1 \right> Y_{l}^{m_2-m_1}(\theta,\varphi),
\label{eq:Gaunt2}
\end{equation}
where the subscript $l=l_{\min},2$ in the summation symbol implies that the summation index $l$ runs in steps of two. The constant $l_{\min}$ is given by~\cite{Weniger-Steinborn-25-149-82}:
\begin{eqnarray}
l_{\min} & = &
\left\{\begin{array}{llll}
\max(|l_1-l_2|,|m_2-m_1|) & \textrm{if} & l_{1}+l_{2}+\max(|l_1-l_2|,|m_2-m_1|) & \textrm{is even}\\ [0.3cm]
\max(|l_1-l_2|,|m_2-m_1|)+1 & \textrm{if} & l_{1}+l_{2}+\max(|l_1-l_2|,|m_2-m_1|) & \textrm{is odd}.
  \end{array}
\right.
\label{EQLMIN}
\end{eqnarray}

A useful property of spherical harmonics is given by:
\begin{equation}
Y_l^m(\theta, \varphi) = (-1)^m \left[Y_l^{-m}(\theta,\varphi)\right]^*.
\label{EQSPHPROP}
\end{equation}

The orthogonality relations of spherical harmonics are defined by:
\begin{eqnarray}
\int_{0}^{\pi} \int_{0}^{2\,\pi}  [Y_{l_1}^{m_1}(\theta,\varphi)]^* Y_{l_2}^{m_2}(\theta,\varphi)\, \sin(\theta) \, d\theta \, d\varphi & = & \delta_{l_1l_2} \delta_{m_1 m_2}
\nonumber\\
\displaystyle \int_{0}^{\pi} \int_{0}^{2\,\pi}  Y_{l}^{m}(\theta,\varphi) \, \sin(\theta) \, d\theta \, d\varphi & = & \delta_{l0} \delta_{m0},
\label{EQORTHOYLM1}
\end{eqnarray}
where $\delta$ stands for the Dirac delta function.

A given function $f(\vec{r})$ and its Fourier transform $\bar{f}(\vec{k})$ are connected by the symmetric relationships:
\begin{equation}
\bar{f}(\vec{k}) = (2\pi)^{-3/2} \int_{\vec{r}} e^{~i \vec{k} \cdot \vec{r}} \; f(\vec{r}) \; d\vec{r} \qquad \textrm{and} \qquad f(\vec{r}) = (2\pi)^{-3/2} \int_{\vec{k}} e^{-i \vec{k} \cdot \vec{r}} \; \bar{f}(\vec{k}) \; d\vec{k}.
\label{eq:TFDef}
\end{equation}

The Fourier integral representation of the Coulomb operator is given by~\cite{Gelfand-Shilov-64}:
\begin{equation}
\frac{1}{\left|\vec{r}\right|} \,=\, \frac{1}{2 \, \pi^2} \, \int_{\vec{k}} \,
\frac{e^{-i \, \vec{k} \,\cdot\, \vec{r}}}{k^2} \,d\vec{k}.
\label{EQFTCOULOMBOP}
\end{equation}

The cartesian coordinates of a vector $\vec{r}$ can be expressed in spherical polar coordinates as a linear combination of spherical harmonics as follows:
\begin{equation}\label{coordinate}
r_{u} \,=\, r \, \sum_{m=-1}^{1} c_{u,m} \, Y_1^m(\theta_{\vec{r}},\phi_{\vec{r}}),
\end{equation}
where $u \in \{ x,y,z \}$ and the coefficients $c_{u,m}$ are given by:
\begin{equation}\label{coeffC}
\left\{
\begin{array}{llllllllll}
c_{x,-1} & = & \sqrt{\frac{2\pi}{3}}, & \quad c_{y,-1} & = & i\, \sqrt{\frac{2\pi}{3}} & \quad\textrm{and}\quad & c_{z,-1} & = & 0 \\[0.25cm]
c_{x, 0} & = & 0, & \quad c_{y, 0} & = & 0 & \quad \textrm{and} \quad & c_{z, 0} & = & \sqrt{\frac{4\pi}{3}}\\[0.25cm]
c_{x, 1} & = & -\sqrt{\frac{2\pi}{3}}, & \quad c_{y, 1} & = & i\, \sqrt{\frac{2\pi}{3}} & \quad\textrm{and}\quad & c_{z, 1} & = & 0.
\end{array}
\right.
\end{equation}

The Pochhammer symbol $(\alpha)_{n}$ is defined by:
\begin{equation}
(\alpha)_n \,=\,
\left\{
  \begin{array}{l}
 (\alpha)_n \,= \, 1 \qquad \textrm{if} \qquad n = 0\\ [0.25cm]
 (\alpha)_n \, = \, \alpha \,(\alpha+1)\,(\alpha+2) \ldots (\alpha+n-1) \, = \,
\displaystyle \frac{\Gamma(\alpha+n)}{\Gamma(\alpha)}
\qquad \textrm{if} \qquad n \leq -\alpha\\ [0.25cm]
 (\alpha)_n \, = \, 0 \qquad \textrm{if} \qquad n \geq -\alpha+1,
  \end{array}
\right.
\label{EQFHS}
\end{equation}where $\Gamma$ stands for the Gamma function. For $n \in \mathbb{N}$:
\begin{equation}
\Gamma(n+1) \,=\,  n! \qquad \textrm{and} \qquad
\Gamma\left(\displaystyle  n+\frac{1}{2}\right) \,=\,  \displaystyle \frac{(2n)!}{2^{2n}\, n!} \,\sqrt{\pi}.
\label{EQGamma}
\end{equation}

\section{Nuclear shielding tensor integrals}

In the presence of an external uniform magnetic field $\vec{B}_{0}$, the electronic non-relativistic Hamiltonian for a system of $n$ electrons and $N$ nuclei is given as a summation over all $n$ electrons of (in atomic units):
\begin{equation}
{\cal H} \,=\,  \sum_{j=1}^{n} \left[\frac{1}{2}\, {\vec{p}_{j}}^{\,2} + \sum_{K=1}^{N} \frac{Z_{K}}{r_{jK}}\right] + \sum_{j=1}^{n}  \sum_{k<j}^{n} \frac{1}{r_{jk}},
\label{EQELECHAMILTON}
\end{equation}
where the electron momentum $\vec{p}_{j}$ is given by:
\begin{eqnarray}
\vec{p}_{j} = \left[- i \, \vec{\nabla}_{j} + e \, \vec{A}(j) \right] \qquad \textrm{where} \qquad \vec{A}(j)= \frac{1}{2} \left(\vec{B}_0 \times \vec{r}_{j0}\right) + \frac{\mu_{0}}{4\, \pi} \sum_{K=1}^{N} \frac{\vec{\mu}_K \times \vec{r}_{jK}}{r_{jK}^3},
\label{EQIMPULVECPOTENTIAL}
\end{eqnarray}
where $\vec{A}(j)$ stands for the vector potential induced by the nuclear moments $\vec{\mu}_{K}$ and the external uniform magnetic field $\vec{B}_{0}$, $\mu_{0}$ stands for dielectric permittivity, $Z_{K}$ is the atomic number of nucleus $K$, $\vec{r}_{j0}$ is the vector distance to the arbitrary gauge origin and where $\vec{r}_{jK} = \vec{r}_{j}-\vec{R}_{K}$, $\vec{r}_{jk}= \vec{r}_{j}-\vec{r}_{k}$. Here $\vec{r}_{j}$ represents the vector position of the electron $j$, $\vec{r}_{k}$ represents the vector position of the electron $k$, and $\vec{R}_{K}$ is the vector position of the nucleus $K$.

Molecular magnetic properties appear as second order perturbative energy corrections. These properties  may be expressed as derivatives of the molecular energy with respect to the nuclear dipole moment and the external field. In the case of nuclear magnetic shielding tensor, the expression is given by:
\begin{equation}
\sigma_{\alpha\beta}^N = \left[ \frac{\partial^2 \left\langle 0 \left| \mathcal{H}(\{\vec\mu\}, \vec B_0) \right| 0 \right\rangle}{\partial B_{0, \alpha} \partial\mu_{N, \beta}} \right]_{\{\vec\mu = \vec 0\},~\vec B_0 = \vec 0},
\label{eq:theory2tensor}
\end{equation}
where $\{\vec\mu\}$ stands for the nuclear magnetic moments, $\vec{B}_0$ is the external magnetic field, and $\mathcal{H}(\{\vec\mu\}, \vec B_0)=\mathcal{H}$ is the total electronic hamiltonian in the presence of the magnetic perturbations. $|0>$ is the ground state wave function. The parameters $\alpha$ and $\beta$ stand for the cartesian coordinates.

A coupled perturbed Hartree-Fock (CPHF) treatment of the equation~(\ref{eq:theory2tensor}) leads to a more explicit expression of the nuclear magnetic shielding tensor~\cite{Ditchfield-56-5688-72, Ditchfield-27-789-74, Stevens-Lipscomb-40-2238-64}:
\begin{equation}
\sigma_{\alpha \beta}^N  = \mathrm{Tr} \left[ P^{(0)^{T}} \cdot h^{(2B\mu_{N,\, \alpha \beta})} + P^{(1B_\alpha)^{T}} \cdot h^{(1\mu_{N,\, \beta})} \right],
\label{eq:theory5}
\end{equation}
where $P^{(0)^{T}}$ and $P^{(1B_\alpha)^{T}}$ are the transpose density matrix of zero order and first order with respect to the external magnetic field. $h^{(1\mu_{N,\, \beta})}$ is the core hamiltonian matrix of the first order with respect to nuclear dipole moment. $h^{(2B\mu_{N,\,\alpha \beta})}$ is the second order one-electron hamiltonian matrix with respect to $B_\alpha$ and $\mu_\beta$. The notation $\mathrm{Tr}$ stands for the trace of the matrix.

Using GIAO, core hamiltonian terms of second orders have the following expressions:
\begin{eqnarray}
h^{(2B\mu_{N,\, \alpha \beta})}_{\mu \nu} & = & \left[ \left( \frac{\partial^2 h}{\partial B_\alpha \partial \mu_{N,\, \beta} } \right)_{\vec{B} = \vec{0},\,\lbrace\vec{\mu}=\vec{0}\rbrace} \right]_{\mu \nu}
\nonumber \\ & = & \frac{i}{2 \, c^2} \; \left[ \Sigma^\alpha_{\mu \nu} \; \left \langle \chi_\mu \left| \frac{ \vec{u}_\beta \cdot \vec{L}_N }{r_{jN}^3} \right| \chi_\nu \right\rangle + \frac{1}{2} \; \left\langle \chi_\mu \left|
\Pi^\alpha_{\mu \nu} \;\frac{\vec{u}_\beta \cdot \vec{L}_N }{r_{jN}^3} \right| \chi_\nu \right\rangle \right]
\nonumber \\ & + & \frac{1}{2c^2} \left\langle \chi_\mu \left|
\frac{(\vec{u}_\alpha \times \vec{r}_{j\nu} )\cdot(\vec{u}_\beta \times \vec{r}_{jN})}{r_{jN}^3} \right| \chi_\nu \right\rangle,
\label{eq:theory7}
\end{eqnarray}
where:
$$
\Sigma^\alpha_{\mu \nu} = \left| \vec{R}_{\mu} \times \vec{R}_{\nu} \right|_\alpha \qquad \textrm{and} \qquad \Pi^\alpha_{\mu \nu} = \left| (\vec{R}_{\mu}-\vec{R}_{\nu}) \times (\vec{r}_{j\mu}+\vec{r}_{j\nu}) \right|_\alpha,
$$
where $\vec{L}_N = -i \,\vec{r}_{jN} \times \vec{\nabla}$ is the angular momentum operator. Here $\vec{u}_{\alpha}$ and $\vec{u}_{\beta}$ are unitary vectors of the cartesian referential.

In~\cite{Safouhi33}, we have developed compact analytical formuale for the one electron three-center nuclear shielding tensor integrals, involved in~(\ref{eq:theory7}) and which are given by:
\begin{equation}
\left\langle \chi_\mu \left| \frac{(\vec{u}_\alpha \times \vec{r}_{j\nu} )\cdot(\vec{u}_\beta \times \vec{r}_{jN})}{r_{jN}^3} \right| \chi_\nu \right\rangle \,=\, \left\langle \chi_\mu \left| \frac{\vec{r}_{j\nu} \cdot \vec{r}_{jN} \, \delta_{\alpha\beta} - r_{jN,\beta}\, r_{j\nu,\alpha}}{r_{jN}^3} \right| \chi_\nu \right\rangle.
\label{eq:int1}
\end{equation}

The main challenge for their analytical development arises from the presence of $\displaystyle \frac{1}{r_{jN}^3}$ in the operator.
In~\cite{Safouhi33}, we used properties of unormalized STFs to express the three-center integral as a linear combination of integrals of the form~:
\begin{equation}
\left\langle \chi_\mu \left | \frac{Y_1^M(\theta_{\vec{r}_{jN}},\varphi_{\vec{r}_{jN}})}{r_{jN}^2} \right| \chi_\nu \right\rangle \qquad \textrm{with} \qquad M = -1, 0, 1.
\label{eq:int5}
\end{equation}

Then, we have expressed the above integral as a linear combination of integrals over $B$ functions using~\eqref{eq:STOonB}, which enables the use of the Fourier transform formalism. We have derived the Fourier transform of the operator in~\eqref{eq:int1} which is given by:
\begin{equation}
\overline{\left(\dfrac{Y_{1}^{M}(\theta_{\vec{r}_{jN}}, \phi_{\vec{r}_{jN}})}{r_{jN}^{2}} \right)}(\vec{k}) \,=\, -i \, \sqrt{\frac{2}{\pi}} \; \frac{Y_{1}^{M}(\theta_{\vec{k}},\phi_{\vec{k}})}{k}.
\label{EQFTOP4}
\end{equation}

In the present contribution, we investigate one-center integrals over $B$ functions.

\section{One-center nuclear shielding tensor integrals over $B$ functions}
If we let $\vec{r}=\vec{r}_{jN}$, the one-center nuclear shielding tensor integrals over $B$ functions are given by:
\begin{eqnarray}
{\cal I}  & = & \left<B_{n_1,l_1}^{m_1}(\zeta_1,\vec{r}_{jN}) \left|\, \mathcal{L}(\vec{r}) \,\right| B_{n_2,l_2}^{m_2}(\zeta_2,\vec{r}_{jN}) \right>_{\vec{r}}
\nonumber\\& = & \int_{\vec{r}} \left[B_{n_1,l_1}^{m_1}(\zeta_1,\vec{r})\right]^* \, \mathcal{L}(\vec{r}) \; B_{n_2,l_2}^{m_2}(\zeta_2,\vec{r}) \, d\vec{r},
\label{EQONECENTINT3a}
\end{eqnarray}
where the operator $\mathcal{L}(\vec{r})$ is given by:
\begin{eqnarray}
\mathcal{L}(\vec{r})  & = & \frac{\vec{r} \cdot \vec{r} \; \delta_{\alpha\beta} - r_{\beta}\, r_{\alpha}}{r^3}
\nonumber\\
& = & \left\{
\begin{array}{lllll}
\dfrac{r^2 - r_{\alpha} \, r_{\alpha}}{r^3}  & = & \dfrac{1}{r} + r_{\alpha}\, \dfrac{\partial}{\partial r_{\alpha}} \Big(\dfrac{1}{r} \Big) \,=\, \dfrac{\partial}{\partial r_{\alpha}} \left(r_{\alpha}\,  \dfrac{1}{r} \right)
& \quad \textrm{if} \quad & \alpha = \beta\\[0.35cm]
\dfrac{- r_{\beta}\, r_{\alpha}}{r^3}   & = &  r_{\beta}\; \dfrac{\partial}{\partial r_{\alpha}}\left(\dfrac{1}{r}\right) & \quad \textrm{if} \quad & \alpha \neq \beta.
\end{array}\right.
\label{EQOPERL}
\end{eqnarray}

By introducing the Fourier transform of the operator $\mathcal{L}(\vec{r})$ given by:
\begin{equation}
\mathcal{L}(\vec{r}) \,=\, (2\, \pi)^{-3/2} \int_{\vec{k}} \overline{\mathcal{L}}(\vec{k})  \, e^{-i \, \vec{k} \,\cdot\, \vec{r}} \, d\vec{k},
\label{EQFTOP}
\end{equation}
in the integral~\eqref{EQONECENTINT3a}, we obtain:
\begin{eqnarray}
{\cal I}  & = & (2\, \pi)^{-3/2} \int_{\vec{k}} \overline{\mathcal{L}}(\vec{k})
\left[\int_{\vec{r}} \left[B_{n_1,l_1}^{m_1}(\zeta_1,\vec{r})\right]^* \, e^{-i \, \vec{k} \,\cdot\, \vec{r}} \; B_{n_2,l_2}^{m_2}(\zeta_2,\vec{r}) \, d\vec{r} \right]\, d\vec{k}
\nonumber\\& = & (2\, \pi)^{-3/2} \int_{\vec{k}} \overline{\mathcal{L}}(\vec{k}) \, \left<B_{n_1,l_1}^{m_1}(\zeta_1,\vec{r}) \left| e^{-i \, \vec{k} \,\cdot\, \vec{r}} \, \,\right| B_{n_2,l_2}^{m_2}(\zeta_2,\vec{r})\right> _{\vec{r}} \, d\vec{k}.
\label{EQONECENTINT3}
\end{eqnarray}

In~\eqref{EQFTOP}, $\overline{\mathcal{L}}(\vec{k})$ stands for the Fourier transfom of $\mathcal{L}(\vec{r})$.

To analytically develop the integral in~\eqref{EQONECENTINT3} using the Fourier transform formalism~\cite{Trivedi-Steinborn-27-670-83, Grotendorst-Steinborn-38-3875-88}, we would need to derive an analytical expression for $\overline{\mathcal{L}}(\vec{k})$ the Fourier transform of the operator $\mathcal{L}(\vec{r})$~\eqref{EQOPERL}.

Let us first start with the case where $\alpha = \beta$. We have~:
\begin{eqnarray}
\overline{\mathcal{L}}(\vec{k}) & = & \overline{\left[\dfrac{\partial}{\partial r_{\alpha}}
\Big(r_{\alpha}\, \dfrac{1}{r}\Big)\right] } (\vec{k})\nonumber\\
& = &  (2 \pi)^{-3/2} \int_{\vec{r}} \, \frac{\partial}{\partial r_{\alpha}}\left(r_{\alpha}\,  \frac{1}{r}\right) \, e^{- i\,\vec{k} \cdot \vec{r}}\, d\vec{r}.
\label{EQFOPERL1}
\end{eqnarray}
Integration by parts, leads to:
\begin{eqnarray}
\overline{\mathcal{L}}(\vec{k})  \,=\,   (2\pi)^{-3/2} \int_{\vec{r}} i\, k_{\alpha}\, r_{\alpha}\,  \frac{1}{r}\, e^{- i \,\vec{k}\cdot \vec{r}} \, d\vec{r}.
\label{EQFOPERL21}
\end{eqnarray}
Using the fact that $r_{\alpha}\,  e^{- i \,\vec{k}\cdot \vec{r}}  \,=\, \dfrac{\partial}{\partial k_{\alpha}}\left[e^{- i\,\vec{k}\cdot \vec{r}}\right]$, we obtain:
\begin{eqnarray}
\overline{\mathcal{L}}(\vec{k}) & = &  - \, k_{\alpha}\, (2\pi)^{-3/2} \int_{\vec{r}}\, \frac{1}{r}\,
\frac{\partial}{\partial k_{\alpha}}\left[e^{- i\,\vec{k}\cdot \vec{r}}\right]\, d\vec{r}
\nonumber\\
& = &  -\, k_{\alpha}\,  \frac{\partial}{\partial k_{\alpha}}\left[ \overline{\left(\frac{1}{r}\right)}(\vec{k})\right].
\label{EQFOPERL22}
\end{eqnarray}

Using the Fourier transform of the Coulomb operator which given by~\eqref{EQFTCOULOMBOP}, we obtain:
\begin{eqnarray}
\overline{\mathcal{L}}(\vec{k}) & = &  -\,  k_{\alpha}\,  \frac{\partial}{\partial k_{\alpha}}\left[ \sqrt{\frac{2}{\pi}}\,  \frac{1}{k^2} \right]
\nonumber\\
& = & \sqrt{\frac{2}{\pi}} \, \frac{2\, k_{\alpha}^2}{k^4}.
\label{EQFOPERL2}
\end{eqnarray}

In the case  where $\alpha$ and $\beta$ do not represent the same Cartesian coordinate, that is $\alpha \neq \beta$, we have:
\begin{eqnarray}
\overline{\mathcal{L}}(\vec{k}) & = & \overline{\left[r_{\beta}\; \dfrac{\partial}{\partial r_{\alpha}}\left(\dfrac{1}{r}\right)\right] } (\vec{k})\nonumber\\
& = &  (2 \pi)^{-3/2} \int_{\vec{r}} r_{\beta}\; \dfrac{\partial}{\partial r_{\alpha}}\left(\dfrac{1}{r}\right)\, e^{- i\,\vec{k} \cdot \vec{r}}\, d\vec{r}.
\label{EQFOPERL3}
\end{eqnarray}
Integration by parts, again, leads to:
\begin{eqnarray}
\overline{\mathcal{L}}(\vec{k}) & = &  -(2\pi)^{-3/2} \int_{\vec{r}} \frac{1}{r} \, \frac{\partial}{\partial r_{\alpha}}\left[r_{\beta}\, e^{- i\,\vec{k}\cdot \vec{r}}\right]\, d\vec{r}
\nonumber\\
& = & -(2\pi)^{-3/2} \int_{r_{\nu}}\int_{r_{\beta}} r_{\beta}\, \left[\int_{r_{\alpha}}  \frac{1}{r} \frac{\partial}{\partial r_{\alpha}}\left[e^{- i\,\vec{k}\cdot \vec{r}}\right] \, dr_{\alpha}\right]\, dr_{\beta}\, dr_{\nu}
\nonumber\\
& = & -(2\pi)^{-3/2} \int_{r_{\nu}}\int_{r_{\beta}} r_{\beta}\, \left[\int_{r_{\alpha}}  \frac{1}{r} \left[- i\, k_{\alpha}\, e^{- i\,\vec{k}\cdot \vec{r}}\right]\, dr_{\alpha}\right] \, dr_{\beta}\, dr_{\nu}
\nonumber\\
& = & k_{\alpha}\, (2\pi)^{-3/2} \int_{r_{\nu}}\int_{r_{\alpha}}\left[\int_{r_{\beta}}  \frac{1}{r}  \left[i\, r_{\beta}\, e^{- i\,\vec{k}\cdot \vec{r}}\right]\, dr_{\beta}\right] \, dr_{\alpha}\, dr_{\nu},
\label{EQFOPERL41}
\end{eqnarray}
where $r_\alpha$, $r_\beta$ and $r_\nu$ represent the cartesian components of the vector $\vec{r}$.

Using again the fact that $r_{\beta}\,  e^{- i \,\vec{k}\cdot \vec{r}}  \,=\, \dfrac{\partial}{\partial k_{beta}}\left[e^{- i\,\vec{k}\cdot \vec{r}}\right]$, we obtain:
\begin{eqnarray}
\overline{\mathcal{L}}(\vec{k})  & = & -k_{\alpha}\, (2\pi)^{-3/2} \int_{r_{\nu}}\int_{r_{\alpha}}\int_{r_{\beta}}  \frac{1}{r} \left[\frac{\partial}{\partial k_{\beta}}\, e^{- i\,\vec{k}\cdot \vec{r}}\right]\, dr_{\beta}\, dr_{\alpha}\, dr_{\nu}
\nonumber\\
& = & -k_{\alpha}\,  \frac{\partial}{\partial k_{\beta}} \, \left[
(2\pi)^{-3/2} \int_{\vec{r}} \frac{1}{r} \, e^{- i\,\vec{k}\cdot \vec{r}}\, d\vec{r}\right]
\nonumber\\
& = & -k_{\alpha}\, \frac{\partial}{\partial k_{\beta}}\,  \left[ \overline{\left(\frac{1}{r}\right)}(\vec{k})\right].
\label{EQFOPERL42}
\end{eqnarray}

From this it follows that:
\begin{eqnarray}
\overline{\mathcal{L}}(\vec{k}) & = &  -k_{\alpha}\, \frac{\partial}{\partial k_{\beta}}\,  \left[ \sqrt{\frac{2}{\pi}}\,  \frac{1}{k^2} \right]
\nonumber\\
& = & \sqrt{\frac{2}{\pi}}\,  \frac{2\, k_{\alpha}\,  k_{\beta}}{k^4}.
\label{EQFOPERL4}
\end{eqnarray}

The Fourier transform $\overline{\mathcal{L}}$ of the operator $\mathcal{L}$ is given by:
\begin{equation}
\overline{\mathcal{L}}(\vec{k}) \,=\,  \sqrt{\frac{2}{\pi}}\,  \frac{2\, k_{\alpha}\,  k_{\beta}}{k^4} \qquad \textrm{for} \qquad
\alpha,\; \beta \,\in\, \{x, y, z\}.
\label{EQFOPERL}
\end{equation}

\section{Fourier transform formalism and the analytical development}

Substituting~\eqref{EQFOPERL} in~\eqref{EQONECENTINT3}, we obtain:
\begin{eqnarray}
{\cal I}  & = & (2\, \pi)^{-3/2} \int_{\vec{k}} \sqrt{\frac{2}{\pi}}\,  \frac{2\, k_{\alpha}\,  k_{\beta}}{k^4}\, \left<B_{n_1,l_1}^{m_1}(\zeta_1,\vec{r}) \left| e^{-i \, \vec{k} \,\cdot\, \vec{r}} \, \,\right| B_{n_2,l_2}^{m_2}(\zeta_2,\vec{r})\right> _{\vec{r}} \, d\vec{k}.
\label{EQONECENTINT31}
\end{eqnarray}

Let us now consider the term ${\cal T} = \left<B_{n_1l_1}^{m_1}(\zeta_{1},\vec{r})\left|
e^{-i \vec{k}\,\cdot\,\vec{r}}\right| B_{n_2l_2}^{m_2}(\zeta_{2},\vec{r})\right>_{\vec{r}}$ involved in the above equation~\eqref{EQONECENTINT31}.

In the integral ${\cal T}$, the two $B$ functions are centered at the same point and the radial part of their product is given by~\cite{Safouhi11}:
\begin{eqnarray}
{\cal T} & = &  \frac{\sqrt{\pi} \, \zeta_1^{l_1} \, \zeta_2^{l_2}\, \zeta_s^{l_1+l_2-1}} {2^{2\,n_1+l_1+2\,n_2+l_2+1} \, (n_1+l_1)! \, (n_2+l_2)!}
\sum_{l=l_{min},2}^{l_{1}+l_{2}} \frac{(-i)^l }{(2\, \zeta_s)^{l}}\, \left<l_1m_1|l_2m_2|lm_1-m_2\right>
\nonumber\\ & \times & \sum_{\tau=2}^{n_1+n_2} \sum_{\varsigma=\tau_1}^{\tau_2}
\frac{2^{\tau}\, (2n_1-\varsigma-1)!\,(2n_2-\tau+\varsigma-1)!\,\zeta_1^{\varsigma-1} \zeta_2^{\tau-\varsigma-1} \zeta_s^{\tau}\, \Gamma(\tau+l_1+l_2+l+1)}
{(\varsigma-1)!\,(n_1-\varsigma)!\,(\tau-\varsigma-1)!\,(n_2-\tau+\varsigma)!\, \Gamma(l+\frac{3}{2})}
\nonumber\\ & \times  & \sum_{r=0}^{\eta^{\prime}} \frac{(-1)^r\, (\frac{\eta}{2})_r \, (\frac{\eta+1}{2})_r }{(l+\frac{3}{2})_r \, r! \, \zeta_s^{2r}}
\, \frac{k^{l+2r}}{\left(\zeta_s^2 + k^2\right)^{\tau+l_1+l_2}} \, [Y_{l}^{m_1-m_2}(\theta_{\vec{k}},\varphi_{\vec{k}})]^{*},
\label{EQONECENTINT12}
\end{eqnarray}
where $\tau_1 = \max (1,\tau-n_2)$, $\tau_2 = \min (n_1,\tau-1)$, $\zeta_s = \zeta_1 + \zeta_2$, $\eta=l-\tau-l_1-l_2+1$ and $\eta^{\prime} = -\frac{\eta}{2}$ if $\eta$ is even, otherwise $\eta^{\prime} = -\frac{\eta+1}{2}$.

The cartesian coordinates of a vector $\vec{r}$ can be expressed in spherical polar coordinates as a linear combination of spherical harmonics as follows:
\begin{equation}\label{coordinate}
r_{u} \,=\, r \, \sum_{m=-1}^{1} c_{u,m} \, Y_1^m(\theta_{\vec{r}},\phi_{\vec{r}}),
\end{equation}

By using~\eqref{coordinate}, we can express the term $\dfrac{k_{\alpha}\, k_{\beta}}{k^4}$ in the Fourier transfor $\overline{\mathcal{L}}(\vec{k})$ in spherical polar coordinates as a linear combination of spherical harmonics as follows:
\begin{equation}
\frac{k_{\alpha}k_{\beta}}{k^4}
= \frac{1}{k^2}\, \sum_{m'=-1}^{1} \sum_{m''=-1}^{1} c_{\alpha,m'} \,
c_{\beta,m''}  \, Y_1^{m'}(\theta_{\vec{r}},\phi_{\vec{r}}) \,Y_1^{m''}(\theta_{\vec{r}},\phi_{\vec{r}}).
\label{coord11}
\end{equation}

Using~\eqref{EQSPHPROP} along with~\eqref{EQLMIN}, we write the above equation as follows:
\begin{equation}
\frac{k_{\alpha}k_{\beta}}{k^4}
= \frac{1}{k^2}\, \sum_{m'=-1}^{1} \sum_{m''=-1}^{1} \sum_{\lambda=\lambda_{\min},2}^{2}  (-1)^{m'} c_{\alpha,m'} \,
c_{\mu,m''}\, \langle 1 \, m''|1\, m'|\lambda \, m''-m' \rangle\,
Y_{\lambda}^{m''-m'}(\theta_{\vec{k}},\phi_{\vec{k}}),
\label{coord12}
\end{equation}
where:
\begin{equation}
\lambda_{\min} =
\left\{
\begin{array}{llll}
|m''-m'|   & \textrm{if} & |m''-m'| & \textrm{is even}\\ [0.3cm]
|m''-m'|+1 & \textrm{if} & |m''-m'| & \textrm{is odd}.
 \end{array}
\right.
\label{EQLMIN11}
\end{equation}

The integration of the angular parts of equations~\eqref{coord12} and~\eqref{EQONECENTINT12} is given by:
\begin{equation}
\int_{0}^{2\pi} \int_{0}^{\pi} [Y_{l}^{m_1-m_2}(\theta_{\vec{k}},\varphi_{\vec{k}})]^{*} \,  Y_{\lambda}^{m''-m'}(\theta_{\vec{k}},\phi_{\vec{k}})\, \sin(\theta){\rm\,d}\theta{\rm\,d}\varphi \,=\, \delta_{l,\lambda}\, \delta_{m_1-m_2,m''-m'}.
\label{EQANGULAR1}
\end{equation}

Using equations~\eqref{EQONECENTINT12} and~\eqref{coord12} and taking into account the result given by~\eqref{EQANGULAR1}, we obtain the following expression for ${\cal I}$:
\begin{eqnarray}
{\cal I} & = & \frac{\zeta_1^{l_1} \, \zeta_2^{l_2}\, \zeta_s^{l_1+l_2-1}}{\pi^{3/2}\, 2^{2\,n_1+l_1+2\,n_2+l_2+2}\, (n_1+l_1)! \, (n_2+l_2)! }
\sum_{m'=-1}^{1} \sum_{m''=-1}^{1} (-1)^{m'} \, c_{\alpha,m'} \, c_{\beta,m''}
\nonumber \\ & \times  & \sum_{\lambda={\lambda}_{\min,2}}^{2} \sum_{l=l_{min},2}^{l_{1}+l_{2}} \frac{(-i)^l}{\left(2 \zeta_s\right)^{l}}  \, \left< 1 \, m''|1 \, m'|\lambda\, m_{4} - m' \right> \, \left<l_1m_1|l_2m_2|lm_1-m_2\right>
\delta_{l,\lambda}\, \delta_{m_1-m_2,m'' - m'}
\nonumber\\ & \times & \sum_{\tau=2}^{n_1+n_2} \, \sum_{\varsigma=\tau_1}^{\tau_2} \frac{2^{\tau}\, (2n_1-\varsigma-1)!\,(2n_2-\tau+\varsigma-1)!\,\zeta_1^{\varsigma-1}\,\zeta_2^{\tau-\varsigma-1}\,\zeta_s^{\tau}\, \Gamma(\tau+l_1+l_2+l+1)}
{(\varsigma-1)!\,(n_1-\varsigma)!\,(\tau-\varsigma-1)!\,(n_2-\tau+\varsigma)!\, \Gamma(l+\frac{3}{2})}
\nonumber\\ & \times  & \sum_{r=0}^{\eta^{\prime}} \, (-1)^r\, \frac{(\frac{\eta}{2})_r \, (\frac{\eta+1}{2})_r}{(l+\frac{3}{2})_r \, r! \, \zeta_s^{2r}} \, \int_{0}^{\infty}\frac{k^{l+2r}}{\left( \zeta_s^2 + k^2\right)^{l_1+l_2+\tau}}{\rm\,d}k,
\label{EQONECENTINT14}
\end{eqnarray}
which can be simplified to:
\begin{eqnarray}
{\cal I} & = & \frac{\zeta_1^{l_1} \, \zeta_2^{l_2}\, \zeta_s^{l_1+l_2-1}}{\pi^{3/2}\, 2^{2\,n_1+l_1+2\,n_2+l_2+2}\, (n_1+l_1)! \, (n_2+l_2)! }
\sum_{m'=-1}^{1} \sum_{m''=-1}^{1} (-1)^{m'} \, c_{\alpha,m'} \, c_{\beta,m''}
\nonumber \\ & \times  & \sum_{l=l_{min},2}^{2} \frac{(-i)^l}{\left(2 \zeta_s\right)^{l}} \, \left< 1 \, m''|1 \, m'|l\, m_{4} - m' \right> \, \left<l_1m_1|l_2m_2|lm_1-m_2\right> \, \delta_{m_1-m_2,m'' - m'}
\nonumber\\ & \times & \sum_{\tau=2}^{n_1+n_2} \, \sum_{\varsigma=\tau_1}^{\tau_2} \frac{2^{\tau}\, (2n_1-\varsigma-1)!\,(2n_2-\tau+\varsigma-1)!\,\zeta_1^{\varsigma-1}\,\zeta_2^{\tau-\varsigma-1}\,\zeta_s^{\tau}\, \Gamma(\tau+l_1+l_2+l+1)}
{(\varsigma-1)!\,(n_1-\varsigma)!\,(\tau-\varsigma-1)!\,(n_2-\tau+\varsigma)!\, \Gamma(l+\frac{3}{2})}
\nonumber\\ & \times  & \sum_{r=0}^{\eta^{\prime}} \, (-1)^r\, \frac{(\frac{\eta}{2})_r \, (\frac{\eta+1}{2})_r}{(l+\frac{3}{2})_r \, r! \, \zeta_s^{2r}} \, \int_{0}^{\infty}\frac{k^{l+2r}}{\left( \zeta_s^2 + k^2\right)^{l_1+l_2+\tau}}{\rm\,d}k.
\label{EQONECENTINT15}
\end{eqnarray}

Now, let us consider the semi-infinite integrals involved in~\eqref{EQONECENTINT15}, and which will be referred to as $\tilde{\cal I}_{\kappa}(\zeta_s)$:
\begin{equation}
\tilde{\cal I}_{\kappa}(\zeta_s) \,=\, \int_{0}^{\infty}\frac{k^{l+2r}}{\left( \zeta_s^2 + k^2\right)^{l_1+l_2+\tau}}{\rm\,d}k
\qquad \textrm{with} \qquad \kappa=l_1+l_2+\tau.
\end{equation}

In order to analytically develop  the semi-infinite integrals $\tilde{\cal I}_{\kappa}(\zeta_s)$, we follow a similar development that we used in~\cite{Safouhi39} for first order relativistic integrals. We first consider $\tilde{\cal I}_{1}(\zeta_s)$:
\begin{equation}
\tilde{\cal I}_{1}(\zeta_s) \,=\, \int_0^\infty \frac{k^{l+2r}}{\zeta_s^2+k^2}{\rm\,d}k.
\end{equation}

By applying the following operator~:
\begin{equation}
\displaystyle\frac{1}{(-2)^{\kappa-1}\Gamma(\kappa)}\left(\frac{\partial}{\zeta_s\,\partial\zeta_s}\right)^{\kappa-1},
\label{EQOPDIFF1}
\end{equation}
to $\tilde{\cal I}_{1}(\zeta_s)$, we obtain the semi-infinite integrals $\tilde{\cal I}_{\kappa}(\zeta_s)$. In other words:
\begin{equation}
\frac{1}{(-2)^{\kappa-1}\Gamma(\kappa)}\left(\frac{\partial}{\zeta_s\,\partial\zeta_s}\right)^{\kappa-1} \, \tilde{\cal I}_{1}(\zeta_s)
\,=\,
\tilde{\cal I}_{\kappa}(\zeta_s).
\end{equation}

Since $l$ is an even number, the integrand of $\tilde{\cal I}_{1}(\zeta_s)$ is an even function, which will be denoted by:
\begin{equation}
f(z) = \frac{z^{l+2r}}{\zeta_s^2+z^2} \qquad \textrm{with} \qquad z=k+i\,y.
\end{equation}
By considering a positively-oriented circular contour above the real axis with radius $R>\zeta_s$ joined at its two ends by the line along the real axis, and by applying Cauchy's residue theorem and taking the limit as $R\to\infty$, we can write:
\begin{equation}
2\int_0^\infty f(k){\rm\,d}k = 2\pi i\,\underset{z=i\zeta_s}{\rm Res}\,f(z).
\end{equation}
By developing further, we obtain the formula:
\begin{equation}
\int_0^\infty \frac{k^{l+2r}}{\zeta_s^2+k^2}{\rm\,d}k = \frac{\pi \,i^{l+2r}}{2}\zeta_s^{l+2r-1}.
\end{equation}
By applying the operator given by~\eqref{EQOPDIFF1} to both sides of the above equation and simplifying, we obtain:
\begin{equation}
\tilde{\cal I}_{\kappa}(\zeta_s) \,=\, \int_0^\infty \frac{k^{l+2r}}{(\zeta_s^2+k^2)^{l_1+l_2+\tau}}{\rm\,d}k = \frac{\pi\,i^{l+2r}}{2}\frac{(-r-\frac{l-1}{2})_{l_1+l_2+\tau-1}}{\Gamma(l_1+l_2+\tau)}\zeta_s^{l+2r-2l_1-2l_2-2\tau+1}.
\label{eq:tildecal1}
\end{equation}

Therefore, by substituting~\eqref{eq:tildecal1} in~\eqref{EQONECENTINT15} and simplifying terms, we finally obtain:
\begin{eqnarray}
{\cal I} & = & \frac{\zeta_1^{l_1} \, \zeta_2^{l_2}\,\zeta_s^{-l_1-l_2}}{\sqrt{\pi}\, 2^{2\,n_1+l_1+2\,n_2+l_2+3}\, (n_1+l_1)! \, (n_2+l_2)! }
\sum_{m_3=-1}^{1} \sum_{m_4=-1}^{1} (-1)^{m_3} \, c_{\alpha,m_3} \, c_{\beta,m_4}
\nonumber \\ & \times  & \sum_{l=l_{min},2}^{2} 2^{-l} \, \left< 1 \, m_4|1 \, m_3|l\, m_{4} - m_3 \right> \, \left<l_1m_1|l_2m_2|lm_1-m_2\right> \, \delta_{m_1-m_2,m_4 - m_3}
\nonumber\\ & \times & \sum_{\tau=2}^{n_1+n_2} \, \sum_{\varsigma=\tau_1}^{\tau_2} \frac{2^\tau\,\zeta_1^{\varsigma-1}\,\zeta_2^{\tau-\varsigma-1}}{\zeta_s^{\tau}} \frac{(2n_1-\varsigma-1)!\,(2n_2-\tau+\varsigma-1)!\,(\tau+l_1+l_2)_{l+1}}
{(\varsigma-1)!\,(n_1-\varsigma)!\,(\tau-\varsigma-1)!\,(n_2-\tau+\varsigma)!\, \Gamma(l+\frac{3}{2})}
\nonumber\\ & \times  & \sum_{r=0}^{\eta^{\prime}} \, \frac{(\frac{\eta}{2})_r \, (\frac{\eta+1}{2})_r\,(-r-\frac{l-1}{2})_{l_1+l_2+\tau-1}}{(l+\frac{3}{2})_r \, r!}.
\label{EQONECENTINT16}
\end{eqnarray}

\section{Conclusion}
In this paper, we show that the Fourier integral transformation can
be applied for the analytical development of the one-center integrals that appear in the second order non relativistic calculations of the nuclear shielding tensor using $B$ function as a basis set of atomic orbitals. The obtained analytical expressions involve semi-infinite integrals which we solved analytically using Cauchy's residue theorem. This leads to compact formulae which can be computed to machine precision without computational difficulty.

\section{Numerical Tables}

In Tables~\ref{SHIELtensxz} and~\ref{SHIELtensyz}, we present values for the integrals ${\cal I}$ of equation~\eqref{EQONECENTINT16}. Table~\ref{SHIELtensxz}, we have $\alpha = x$ and $\beta = z$, and in  Table~\ref{SHIELtensyz}, we have $\alpha = y$ and $\beta = z$.

For the numerical evaluation of Gaunt coefficients which occur in the complete expressions of the integrals under consideration, we use the subroutine GAUNT.F developed by Weniger et al.~\cite{Weniger-Steinborn-25-149-82}. The spherical harmonics $Y_{l}^{m}(\theta,\varphi)$ are computed using the recurrence formulae presented in~\cite{Weniger-Steinborn-25-149-82}.

In all Tables, the numbers in parentheses represent powers of $10$.

\begin{center}
\begin{table}[!hp]
\caption{Evaluation of ${\cal I}$~\eqref{EQONECENTINT16} for $\alpha=x$ and $\beta=z$.\label{SHIELtensxz}}
\begin{tabular*}{\hsize}{@{\extracolsep{\fill}}ccccccccc} \sphline
~~$n_1$&$l_1$&$m_1$&$\zeta_1$&$n_2$&$l_2$&$m_2$&$\zeta_2$&${\cal I}$\\ \sphline
 2 & 1 & ~0 & 0.125 & 2 & 1 & -1 & 0.125 & ~.300398510436534(-3)\\
 3 & 2 & -1 & 0.125 & 2 & 1 & ~0 & 0.125 & ~.850537627015333(-4)\\
 3 & 2 & ~2 & 0.125 & 2 & 1 & ~1 & 0.125 & -.120284184743371(-3)\\
 3 & 2 & ~2 & 0.125 & 3 & 2 & ~1 & 0.125 & -.323872527763475(-4)\\
 3 & 2 & -1 & 0.125 & 3 & 2 & ~0 & 0.125 & ~.132220405787649(-4)\\
 3 & 2 & -1 & 0.125 & 3 & 2 & -2 & 0.125 & ~.323872527763475(-4)\\
 4 & 3 & ~2 & 0.125 & 3 & 2 & ~1 & 0.125 & -.106263923365582(-4)\\
 4 & 3 & ~3 & 0.125 & 3 & 2 & ~2 & 0.125 & -.130146195155945(-4)\\
 4 & 3 & ~3 & 0.125 & 4 & 3 & ~2 & 0.125 & -.297056955276114(-5)\\
 5 & 4 & ~2 & 0.125 & 4 & 3 & ~1 & 0.125 & -.952021208326728(-6)\\
 5 & 4 & ~4 & 0.125 & 4 & 3 & ~3 & 0.125 & -.130070859331645(-5)\\
 5 & 4 & ~3 & 0.125 & 4 & 3 & ~2 & 0.125 & -.112644668473277(-5)\\
 5 & 4 & ~4 & 0.125 & 4 & 3 & ~3 & 0.125 & -.130070859331645(-5)\\
 6 & 5 & ~2 & 0.125 & 5 & 3 & ~1 & 0.125 & -.196767017372175(-6)\\ \hline
\end{tabular*}
\end{table}

\begin{table}[!hp]
\caption{Evaluation of ${\cal I}$~\eqref{EQONECENTINT16} for $\alpha=y$ and $\beta=z$.\label{SHIELtensyz}}
\begin{tabular*}{\hsize}{@{\extracolsep{\fill}}ccccccccc} \sphline
~~$n_1$&$l_1$&$m_1$&$\zeta_1$&$n_2$&$l_2$&$m_2$&$\zeta_2$&${\cal I}$\\ \sphline
 3 & 1 & ~0 & 0.125 & 2 & 1 & -1 & 0.125 & ~.225413363967049(-3) \\
 4 & 2 & -1 & 0.125 & 2 & 1 & ~0 & 0.125 & -.565646839081594(-4) \\
 4 & 2 & ~2 & 0.125 & 2 & 1 & ~1 & 0.125 & -.799945431342662(-4) \\
 4 & 2 & ~2 & 0.125 & 3 & 2 & ~1 & 0.125 & -.234135564517191(-4) \\
 4 & 2 & -1 & 0.125 & 3 & 2 & ~0 & 0.125 & -.955854439509345(-5) \\
 4 & 2 & -1 & 0.125 & 3 & 2 & -2 & 0.125 & ~.234135564517191(-4) \\
 5 & 3 & ~2 & 0.125 & 3 & 2 & ~1 & 0.125 & -.718385939252909(-5) \\
 5 & 3 & ~3 & 0.125 & 3 & 2 & ~2 & 0.125 & -.879839494779829(-5) \\
 5 & 3 & ~3 & 0.125 & 4 & 3 & ~2 & 0.125 & -.214692474189422(-5) \\
 5 & 4 & ~2 & 0.125 & 4 & 3 & ~1 & 0.125 & -.952021208326728(-6) \\
 5 & 4 & ~3 & 0.250 & 4 & 3 & ~2 & 0.125 & -.458285620562908(-6) \\
 5 & 4 & ~4 & 0.125 & 4 & 3 & ~3 & 0.125 & -.130070859331645(-5) \\
 6 & 5 & ~2 & 0.125 & 5 & 2 & ~1 & 0.125 & -.384684858951031(-6) \\
 6 & 5 & ~2 & 0.125 & 5 & 3 & ~1 & 0.125 & -.196767017372175(-6) \\ \hline
\end{tabular*}
\end{table}
\end{center}

\clearpage

\end{document}